# AstroTwitter

S. R. Lowe

Jodrell Bank Centre for Astrophysics, The University of Manchester

**Summary**. AstroTwitter aims to make it easy for both professional and amateur telescopes to let the world know what they are observing in real-time.

## 1 Introduction

In 2001 I began my PhD at Jodrell Bank Observatory in Cheshire, UK. Jodrell Bank - part of the University of Manchester - is a working observatory with four radio telescopes on the site. The largest telescope is the Lovell Telescope which, with a diameter of 76m, stands tall above the Cheshire plain. Near the base of the telescope is the observatory's Visitor Centre which was first opened in 1971. At its peak the Visitor Centre attracted over 120,000 visitors per year and even in its current reduced state attracts over 70,000 visitors annually.

Over the years I've spent time in the Visitor Centre helping with events and answering countless astronomical questions from members of the public. As people can see and hear the Lovell telescope slowly tracking the sky and occasionally changing target, one of the common questions I've heard visitors ask is "What is it looking at?" In fact, this question probably arises in the minds of everyone who sees a professional telescope observing the sky. The question is simple to ask but often the answer is not so easy to obtain.

## 2 What is it looking at?

My first attempts to answer the question "what is it looking at?" started modestly. I was building new receivers for a small radio telescope that the public would use to observe the Sun.

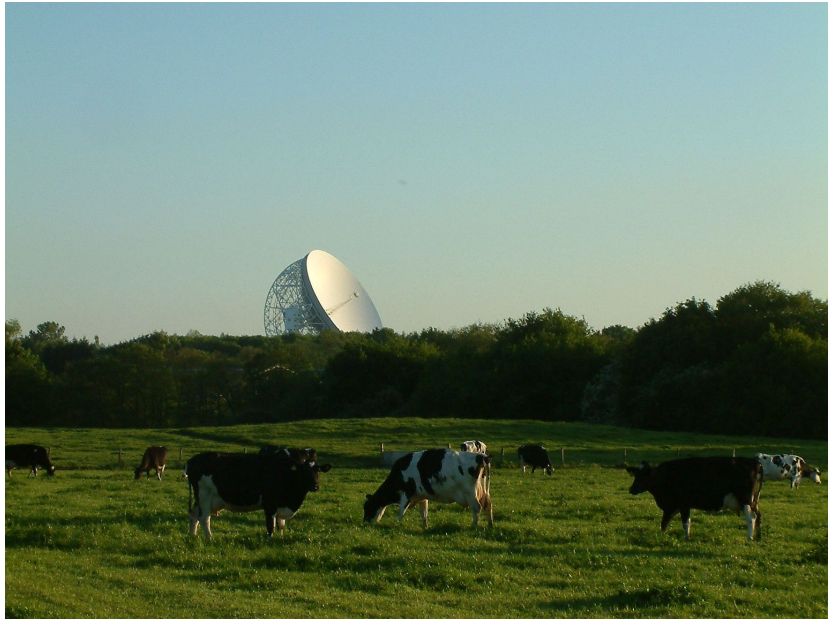

**Fig. 1.** "What is it looking at?" This question is commonly asked about the Lovell Telescope by visitors to Jodrell Bank Observatory.

To help the public steer the telescope, I wrote software which displayed the observing co-ordinates of the telescope in the context of the sky visible from the UK. This included a simple display showing the positions of the telescope, the Sun and the brightest stars. With some minor effort by engineers at the observatory, I was later able to pipe the live observing coordinates of all seven of the Observatory's telescopes into the same software display.

The output of the Jodrell Bank software is displayed on a screen in the Visitor Centre and as an image on a website. This limits the ways in which the information can be re-used or displayed and is totally unusable by the visually impaired. To become more useful to visitors and staff it was necessary to find a better way to share this information.

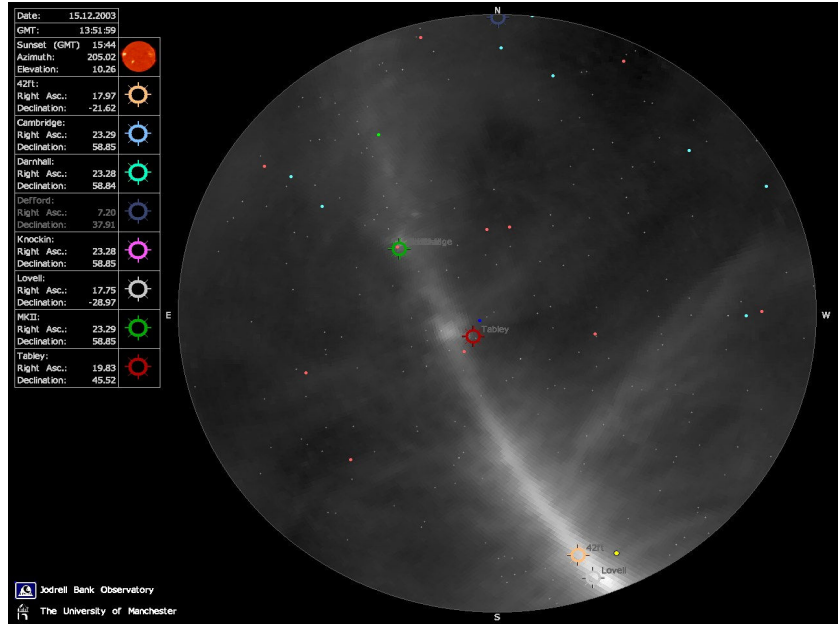

**Fig. 2.** An early incarnation of the software which displays live observing positions for Jodrell Bank's radio telescopes. The telescopes' coordinates are displayed on top of a 408 MHz radio sky (background image: Haslam et al, 1982). Image credit: JBCA, The University of Manchester.

## 3 All A-twitter

In March 2006 a new online service, Twitter, emerged. Twitter (www.twitter.com) is a "service for friends, family, and co-workers to communicate and stay connected through the exchange of quick, frequent answers to one simple question: What are you doing?"

A form of microblogging, Twitter allows updates of up to 140 characters to be sent from a preferred interface; the web, text message (SMS), instant messenger (IM) or your own application. Users of the service sign up to follow other users and receive their updates in most* of these forms too. This gives people fantastic scope to communicate everything from the intricate details of what they are having for lunch to less trivial matters such as news stories, earthquake reports and even summaries of service calls to the Los Angeles Fire Department. It is the mix that makes it so interesting.

Twitter's simple application programming interfaces (APIs) - the ability to get information in and out of the service - made it a natural choice to broaden the usability of the live telescope data. Within minutes it was possible to register an account and start piping the telescope positions and observing targets to it. The hard work of generating valid syndicated feeds (e.g. RSS) and providing output to a variety of platforms was taken care of automatically.

### 3.1 In Space Everyone Can Hear You Tweet

Jodrell Bank's telescopes were amongst the first astronomical users of Twitter and have since been joined by many astronomers, engineers and outreach professionals who use the service to network and solicit answers to questions e.g. "*Does anyone know if Mercury will be visible in Sept. of 09? I'm having trouble determining it.*".

The most successful astronomical feed on Twitter is NASA's Mars Phoenix lander (www.twitter.com/MarsPhoenix). This feed was initially created as a way for those working on the project to stay up-to-date with the preparations for landing during a holiday weekend in the United States. As Mars Phoenix started to make the front pages of newspapers and online news websites, the number of Twitter followers rapidly increased. During the May 25th landing it already had 3000 followers and that had increased to over 35,000 by the end of September 2008 making it one of the most popular feeds on the service. These followers have been kept informed of progress every step of the way and were amongst the first to learn about possible water ice dug up by the lander. They will also be informed when the Martian winter brings an end to the mission; an event that is unlikely to be reported widely in the mainstream media.

The Mars Phoenix team did two things that worked well in the medium. The first was the decision to write in the first person. This made good sense because, with a 140 character limit, every character matters. Using 'I' rather than 'Mars Phoenix' saves precious space but it also gives the spacecraft a personality and encouraged interaction from the public. Rather than ignore the comments and questions they received, the team have actively replied to questions from their followers. All 35,000 followers can see the answers in their own Twitter streams.

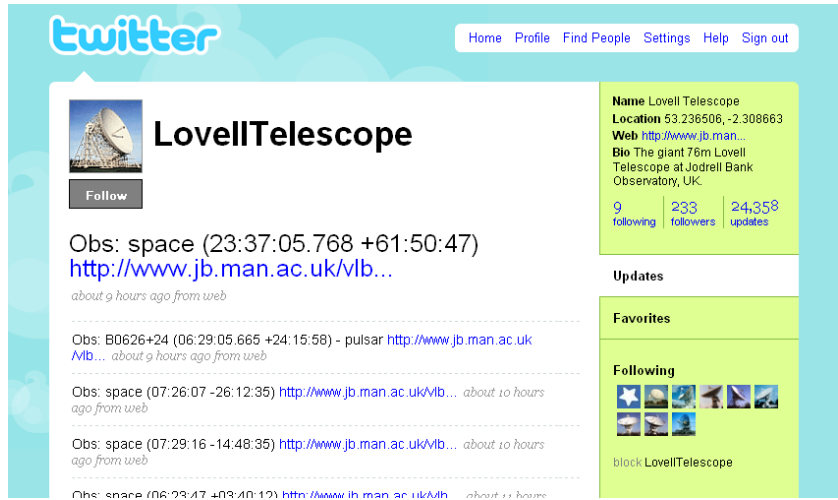

**Fig. 3.** The Lovell Telescope's twitter stream showing recent observations.

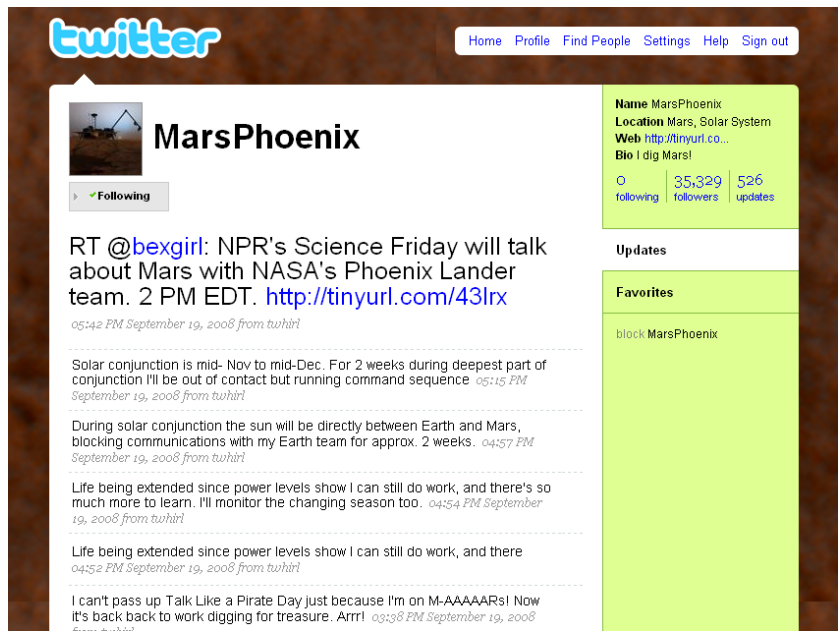

**Fig. 4.** Mars Phoenix's twitter stream.

Since Mars Phoenix, many other active spacecraft, or those under construction, have started twittering. Planck (www.twit-

ter.com/Planck), Lunar Reconnaissance Orbiter/LCROSS (www.twitter.com/LRO_NASA), and the Solar Dynamics Observatory (www.twitter.com/NASA_SDO_HMI) are all reporting the progress of their ground tests in preparation for launches during International Year of Astronomy 2009. The Cassini mission is reporting the results of fly-bys of Saturn's moons and the Hubble Space Telescope links to its latest images, press releases and scheduled observations. As well as following the recipe of Mars Phoenix, some of these missions have even been chatting to each other about their shared experiences and communicating their excitement as they get closer to launch.

**Fig. 5.** ESA's Planck spacecraft regularly 'chats' to other spacecraft.

With a direct line to users, Twitter can also be used to supplement other forms of outreach. The LRO team combined their efforts across several online platforms to collect over 1.55 million names which they will be taking to the Moon. A large number of these names came via campaigns on Twitter and Facebook.

Used well, Twitter gives the public a way to interact with mission teams before, during and after launch in a more immediate and intimate way than was possible previously.

## 4 AstroTwitter

Twitter is a great service but is far from unique or perfect. It is run by a relatively small team and the service suffers occasional outages (colloquially known as the "fail whale" to users of Twitter) as they struggle with the demand. The generic nature of the service is also a limitation for astronomical applications as it is difficult to efficiently communicate everything required to define an observation and then do other, more interesting things when you only have 140 characters.

A dedicated AstroTwitter aims to make it easier to answer the question "What are we looking at?" for many telescopes around the world in a consistent way. This will be achieved by providing a Twitter-like service tuned to the needs of telescopes.

### 4.1 Providing Pipes

In practice, each observatory or telescope (amateur and professional) would create its own account on AstroTwitter. They would be able to define various details about their account such as their username, a short biography or description, their location and an image. With straightforward APIs, it will be easy for those telescopes to pipe their data into the service in the most appropriate way for them. Being dedicated to astronomy, these data can then be provided in a wider variety of specialized output formats than currently provided by Twitter. These could include:
- Webpage
- RSS feeds
- Simple XML feeds
- Google Sky overlays
- Twitter updates
- Website widgets
- VOEvents
- Links to nearby objects (Aladin)
- Links to papers about the object (ADS)
- Links to press release images (VAMP)

It should also be possible to provide some of these outputs for user-defined groups of telescopes.

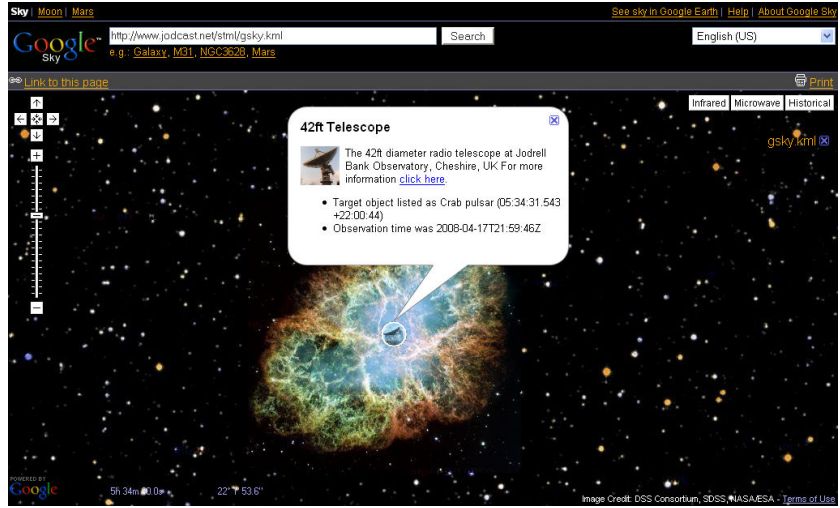

**Fig. 6.** A working example of Jodrell Bank's Google Sky feed showing the *42ft Telescope* observing the Crab pulsar. Image credit: JBCA, DSS Consortium, SDSS, NASA/ESA, Google Sky.

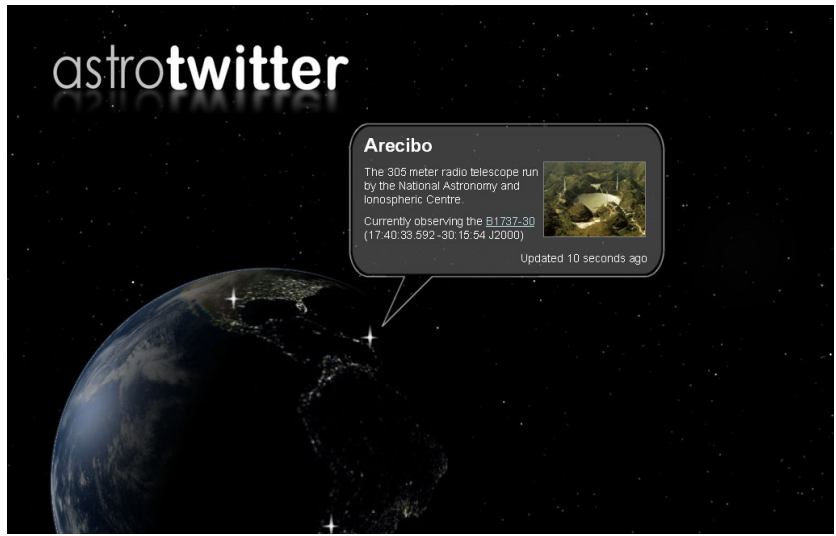

**Fig. 7.** Existing mash-ups such as twittearth.com display tweets on a 3D representation of the Earth. This could be adapted to display the current status of telescopes from the AstroTwitter aggregator.

**4.2 To tweet or not to tweet**

The barriers to the implementation of this idea will be mostly due to technology or data issues at individual observatories; there may be no easy computer access to real-time observing coordinates or telescopes may not be connected to the Internet. To join AstroTwitter some initial effort may be required on the part of each observatory to implement but should require little maintenance once automated.

Even when the technological problems are overcome, there can be times when it may be necessary to keep observing coordinates private. In cases where discovery observations of a potential minor body or extrasolar planet are being taken, public knowledge of the position could lead to another astronomer scooping the discovery announcement. These considerations need not deter an observatory from taking part as the proposed model for AstroTwitter leaves control over the submission of data to each individual observatory. When required, the telescope could simply remain quiet.

Ultimately, the telescope will gain more exposure amongst the public and other astronomers if it becomes part of the community.

**5. Conclusions**

AstroTwitter will allow both professional and amateur observatories to provide feeds and services showing where they are observing in real-time. The hard work of creating multiple output formats will be done by AstroTwitter ensuring that individual observatories can keep up with changing formats and applications for a minimal initial investment of time.

One obvious use of the generated output is to display it in applications such as Google Sky, Microsoft's World Wide Telescope or Stellarium. This will allow the public to see a real-time snap-shot of what the world's observatories are looking at displayed within a realistic representation of the sky. Not only will this answer the original question - what is it looking at? - it could also inspire the public to observe the same objects themselves with the thrill of knowing that some of the biggest telescopes on (or off) the planet were looking at the same thing.

With AstroTwitter it would be possible to watch telescopes chasing gamma-ray burst alerts, follow Very Long Baseline Interferome-try observations or even check if your scheduled observations were being made. With an archive it would be possible to create sky maps showing areas that are observed regularly or those that are relatively neglected. If it is easy to extract information from the system, many more interesting and innovative applications will emerge over time.

## Footnotes

* Receiving messages from Twitter by text message (SMS) was discontinued in many countries during 2008 because of the way in which local phone providers charge. Sending updates *to* Twitter by text message is still possible.